\documentclass[twocolumn,pre,superscriptaddress,preprintnumbers,amsmath,amssymb,longbibliography]{revtex4-1}
\usepackage{graphicx}
\usepackage{color}

\def\gtrless{\raise2.5pt\hbox{$>$}\llap{\lower2.5pt\hbox{$<$}}}
\def\gtrapprox{\raise2.5pt\hbox{$>$}\llap{\lower2.5pt\hbox{$\approx$}}}

\newcommand{\bsq}[1]{\begin{subequations}\label{#1}}
\newcommand{\esq}{\end{subequations}}
\newcommand{\beq}[1]{\begin{equation}\label{#1}}
\newcommand{\eeq}{\end{equation}}
\newcommand{\beqa}[1]{\begin{eqnarray}\label{#1}}
\newcommand{\eeqa}{\end{eqnarray}}

\newcommand{\mb}{\mathbf}

\renewcommand{\rho}{\varrho}
\renewcommand{\epsilon}{\varepsilon}

\begin{document}

\title{Dynamics of active filaments in porous media}

\author{ Zahra Mokhtari}
\author{Annette Zippelius}
\affiliation{University G\"ottingen, D-37077 G\"ottingen, Germany}

\date{\today}



\begin{abstract} The motion of active polymers in a porous medium is
  shown to depend critically on flexibilty, activity and degree of
  polymerization. For given Peclet number, we observe
  a transition from localisation to diffusion as the stiffness of the
  chains is increased. Whereas stiff chains move almost unhindered
  through the porous medium, flexible ones spiral and get stuck. Their
  motion can be accounted for by the model of a continuous time random
  walk with a renewal process corresponding to unspiraling. The
  waiting time distribution is shown to develop heavy tails for
  decreasing stiffness, resulting in subdiffusive and ultimately caged behaviour.
\end{abstract}


\maketitle
Understanding the motion of biological agents in porous media is 
essential for a wide range of biological, medical and industrial
processes. Mucus, a hydrogel which coats the stomach and
other intestinal walls, serves as an important defense against
invading bacteria~\cite{cornick2015roles,laux2005role,cohen199524,celli2009helicobacter}. In
fighting cancer, bacteria are now engineered to sense the porous
environment of a
tumor~\cite{anderson2006environmentally,felfoul2016magneto}.  Other
bacteria such as Myxobacteria glide in the porous environment of soil
and synthesize a number of biomedically useful
chemicals~\cite{dawid2000biology}. Yet another exmaple of active
motion in a porous medium is the crowded world of a cell, in which
active filaments move in a polymeric gel.  Technical applications
involve the motion of bacteria in porous media in the context of oil
recovery~\cite{brown2010microbial}, water purification and
decomposition of contaminants trapped in the
ground~\cite{ginn2002processes,simon2002groundwater}.

It is known that the dynamics of biological agents in porous media is
strongly influenced by their shape, size, and properties of the media
such as the pore sizes~\cite{spagnolie2013locomotion, fu2010low,
  leshansky2009enhanced}. In spite of many experimental studies on the
motion of active elongated agents in porous
media~\cite{croze2011migration,creppy2019effect,brown2016swimming,wolfe1989migration,martinez2014flagellated,
  sosa2017motility,raatz2015swimming}, theoretical approaches are
sparse~\cite{barton1997mathematical,bertrand2018optimized}. In contrast, the flow of passive
polymers in porous media has been studied extensively. Whereas flexible chains
can be represented by a sequence of blobs~\cite{Yamakov1997} whose
size is determined by the cavities of the porous media, the dynamics
of stiff polymers was found to follow the reptation
picture~\cite{Nam2010} with, however, different kinetic exponents.

Bacteria and many other microorganism have an elongated shape and
resist deformations with a finite stiffness. Their flexibility plays
an important role in their dynamics: While stiff active filaments such
as microtubules form large coherently moving bundles~\cite{sanchez2011cilia,lam2016cytoskeletal,inoue2015depletion},
active agents with flexible bodies, such as bacteria, have been found
to form slowly diffusing spirals~\cite{lin2014dynamics}. Analytical
and numerical studies of active filaments have revealed that bundles
and spirals form in different regions of phase space, solely
determined by the activity and stiffness of the
chains~\cite{liverpool2003anomalous,isele2015self,prathyusha2018dynamically}.
Flexibility is even more important in a crowded environment: a large
number of biological agents maximize their transport by deforming
their shape depending on the environment and interaction with other
objects~\cite{menzel2012soft,ohta2009deformable}.  The interest in
active elongated agents is further driven by new developments to
prepare them in
vitro~\cite{schaller2010polar,sumino2012large,kraikivski2006enhanced}
and synthesize them, e.g. by means of bonding several Janus colloidal
spheres through electric fields~\cite{yan2016reconfiguring}, and also
by immersing chains of passive colloidal particles in an active bath,
where directed transport has been
observed~\cite{sasaki2014colloidal}.

In this work we analyse the motion of self propelled filaments in a
porous medium with help of numerical simulations, supported by
analytical arguments. Our main result is a phase diagram with a
diffusive and a localised phase, whose intuitive interpretation is
based on the competition between the activity driving the polymers
through narrow channels and the flexibility favouring a spiraling, or
more generally folded state of the filaments which then are caged in
the porous medium.
Our model consists of three parts: a standard model for a
semiflexible chain, overdamped dynamics of the chain including active
beads and an ensemble of static obstacles.
We consider a wormlike chain, consisting of $M$ active beads of radius $R$ at
positions $\mb{r}_i, i=1,2...M$. The potential energy
$U_{pol}=U_s+U_b+U_e$ of the chain has three contributions: The connectivity of the chain is modelled by springs
with spring constant $K_s$ and
rest length $b$. The bending energy is given by
\begin{equation}
  U_b=\frac{K_a}{2}\sum_{i=2}^{M-1} \big(\theta_{i}-\pi \big)^2,
 \qquad \cos{\theta_{i}}=
\frac{\mb{r}_{i,i-1}\cdot\mb{r}_{i+1,i}}{{r}_{i,i-1}{r}_{i+1,i}}
\end{equation}
with $\mb{r}_{i,i-1}=\mb{r}_i-\mb{r}_{i-1}$ and bending stiffness $K_a$. 
The excluded volume is modeled as a contact potential
\begin{equation}\label{repulsive_potential}
	U_e = \frac{1}{2} K_e \sum_{i\neq j}(r_{i,j}-2R)^2\;\theta(2R-r_{i,j})
\end{equation} 
We are interested in approximately hard particles of fixed bond length
and hence take both, $K_s$ and $K_e$, to be large. To avoid chain
crossings, the bond length $b=2.1 R$ is chosen only slightly larger
than the diameter of the beads. This leaves us with one free
parameter, namely the bond stiffness, $K_a$, or rather the persistence
length $\xi/L=2bK_a/(Lk_BT)$ relative to the contour length $L=Mb$.

The dynamics of the chain is assumed to be overdamped
\begin{equation}\label{dynamics}
  \gamma \dot{\mb{r}}_i=F_{act}\mb{t}_i-\nabla_i U+\boldsymbol{\eta}_i
\end{equation}
The active force has amplitude $F_{act}$ and points along the tangent
of the polymer contour,
$\mb{t}_i=(\mb{r}_{i,i-1}+\mb{r}_{i+1,i})/|\mb{r}_{i,i-1}+\mb{r}_{i+1,i}|$.
The potential $U$ includes $U_{pol}$ as well as the interactions with
the obstacles, yet to be specified. The damping constant is denoted by
$\gamma$ and the random noise is chosen in accordance with the
fluctuation dissipation theorem:
$<\boldsymbol{\eta}_i(t)\cdot\boldsymbol{\eta}_j(t^{\prime})>=
4k_BT\delta_{ij}\gamma\delta(t-t^{\prime})$. It is convenient to work with
dimensionless equations. Hence we measure length in units of particle
radius $R$ energies in units of $k_BT$ and time in units of
$t_0=R^2\gamma/(k_BT)$. In these units the active force is given by
$(F_{act}R)/(k_BT)$ and represents the second control parameter besides
the persistence length. Actually it is more convenient to multiply
this quantity by the dimensionless factor $L^2/(bR)$ in order to
obtain the Peclet number $\rm{Pe}=(F_{act}L^2)/(bk_BT)$, which is defined as the ratio of the convective transport to the diffusive transport. 

So far our model is the same as the one used in
ref.(\cite{isele2015self}).  However, here we are interested in the
dynamics of semiflexible chains in a crowded environment. We introduce
N static obstacle of radius $R_o$ randomly into a two-dimensional
square box of size $\ell$. The interaction between particles and obstacles are taken to be elastic collisons, reversing the normal component of the relative velocity~\cite{Mokhtari2017}. We impose the constraint that obstacles do
not overlap, so that their packing fraction is given by
$\phi=N\pi R_o^2/\ell^2$. Of particular interest is a porous medium with a
typical pore size, which is modelled with help of an additional constraint:
the relative distance between any two obstacles is at
least $2R_o+2.5 R$, allowing single beads to pass in between two
obstacles. The central questions of our study are the following:
{\it Are the polymers free to move through the medium or are they localised?
   How does their dynamics depend on chain length, stiffness and Peclet
  number?}  A similar porous medium has been set up in a recent
experiment~\cite{creppy2019effect}, where the motion of bacteria in
the presence of randomly placed pillars in microfluidic chips has been
studied.

Eq.~\ref{dynamics} was integrated using
HOOMD-blue~\cite{anderson2008general,glaser2015strong} with an
in-house modification to include the active force along the tangent of
the polymer contour. Simulations were performed on graphical
processing units (GPUs). We set $K_s=K_e=500k_BT/R^2$ and $M=30$
unless said otherwise, and explore a large parameter space for $\rm{Pe}$
and $\xi/L$ by varying $F_{act}$ and $K_a$.  Measurements are carried
out after an initial time lapse, and for as long as $10^6 t_0$.

\begin{figure}
\includegraphics[width=0.23\textwidth]{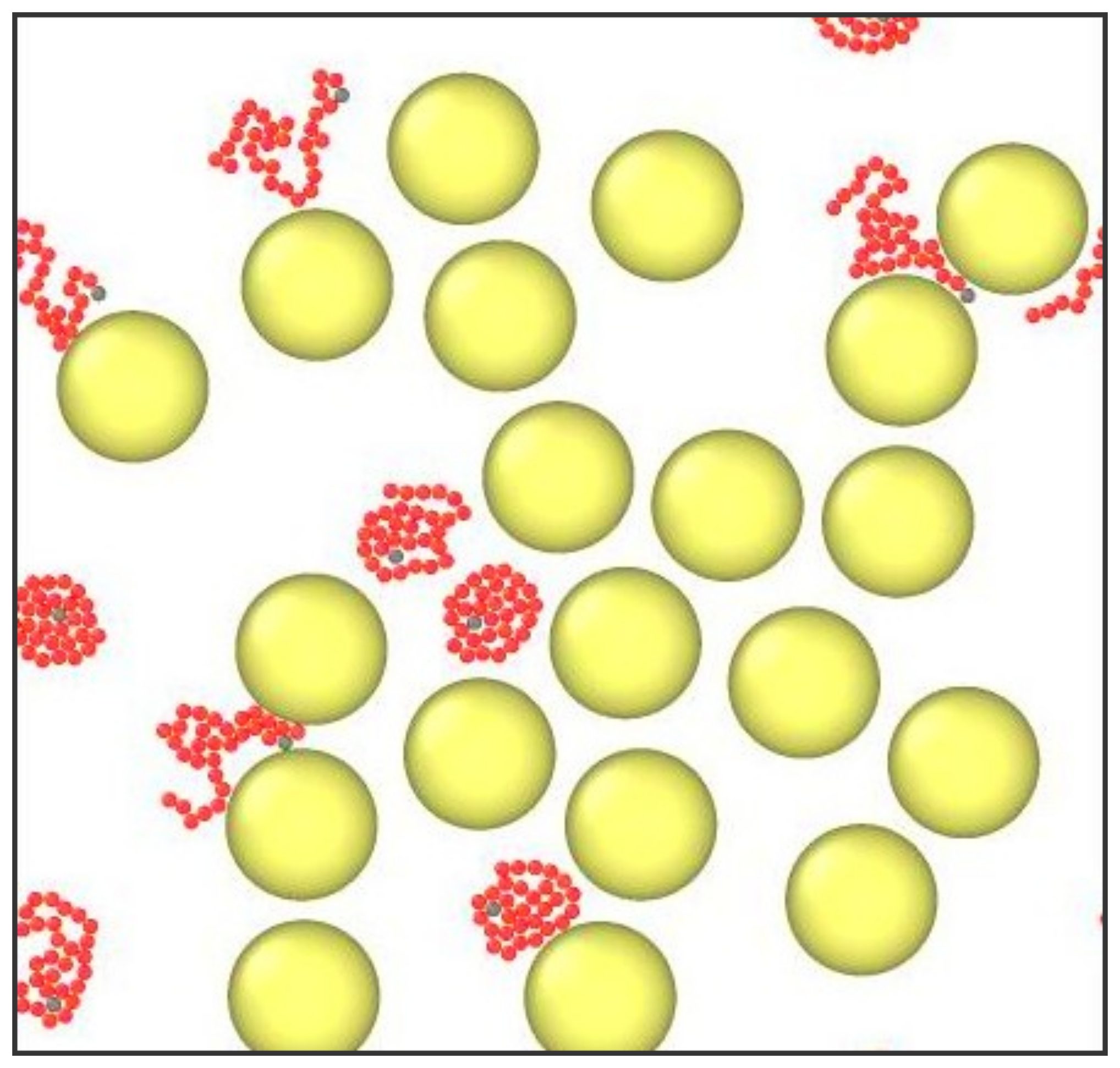}
\includegraphics[width=0.23\textwidth]{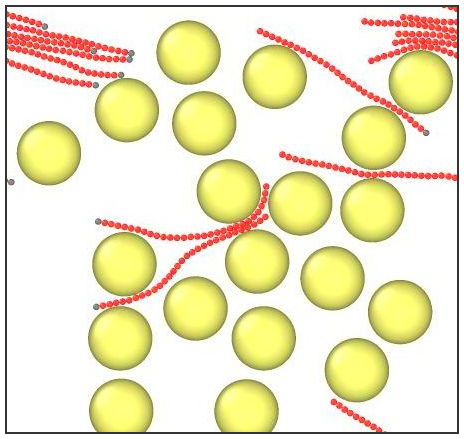}  
\caption{Flexible polymers (left) spiral, sometimes around the obstacles; stiff polymers (right) bend only slightly to pass around an obstacle.}
  \label{spirals}
\end{figure}

The central topic of our paper is the selective localisation of stiff
and flexible polymers in an obstructed medium. Before addressing this
subject in detail, we comment on the diffusion of semiflexible
polymers in an unobstructed environment which has been discussed in
recent literature~\cite{isele2015self,prathyusha2018dynamically}. One
of the most prominent results is the spiraling phase, where flexible
polymers form persistent spirals. We of course also observe these
states, sometimes the flexible polymers spiral around an obstacle,
provided their contour length is larger than the circumference of the
obstacle. These results are to be expected and a sample configuration
is shown in Fig.~\ref{spirals}.
The stiff polymers tend to follow the obstacles only for short
sections of their contour length, compromising the cost of bending
energy and activity $F_{act}L$. This qualitatively different
behaviour is at the heart of the selective transport, discussed in
detail in the following sections.

For a single polymer, obeying the dynamics of Eq.(\ref{dynamics}), we
can compute the centre of mass diffusion (without obstacles), because all interactions are
pairwise and do not contribute to the CM motion:
\begin{equation}
  \dot{\mb{R}}_{CM}=\frac{F_{act}}{\gamma L}\mb{R_{e}}+\frac{1}{M}\sum_{i=1}^M\boldsymbol{\eta}_i
\end{equation}
Here $\mb{R}_e$ denotes the end-to-end vector.
Since the activity points along the contour, the polymer motion is well approximated by ``railway motion''~\cite{isele2015self}, 
allowing for the computation of the end-to end vector within the Kratky-Porod model.
The mean square displacement (MSD) then follows
\begin{align}
  <(\mb{R}(t)&- \mb{R}(0))^2>=4D_tt\nonumber\\
 + &\frac{2\xi F_{act}}{\gamma} f(L/\xi)\big(t+(e^{-D_rt}-1)/D_r\big).
\end{align}
The activity induced rotational diffusion constant is given by
$D_r=F_{act}/(\xi\gamma)$~\cite{isele2015self} and
$f(x)=2(e^{-x}-1+x)/x^2$.  The dynamics is diffusive for short
times with bare diffusion constant $D_t=(k_BT)/(\gamma M)$, ballistic
for intermediate times and diffusive again at long times with,
however, an enhanced diffusion constant
$D_{eff}=D_t+F_{act}\xi f(L/\xi)/(2\gamma)$, due to the activity of the chain.

\begin{figure}[h!]
		\center	
		\includegraphics[width=0.49\textwidth]{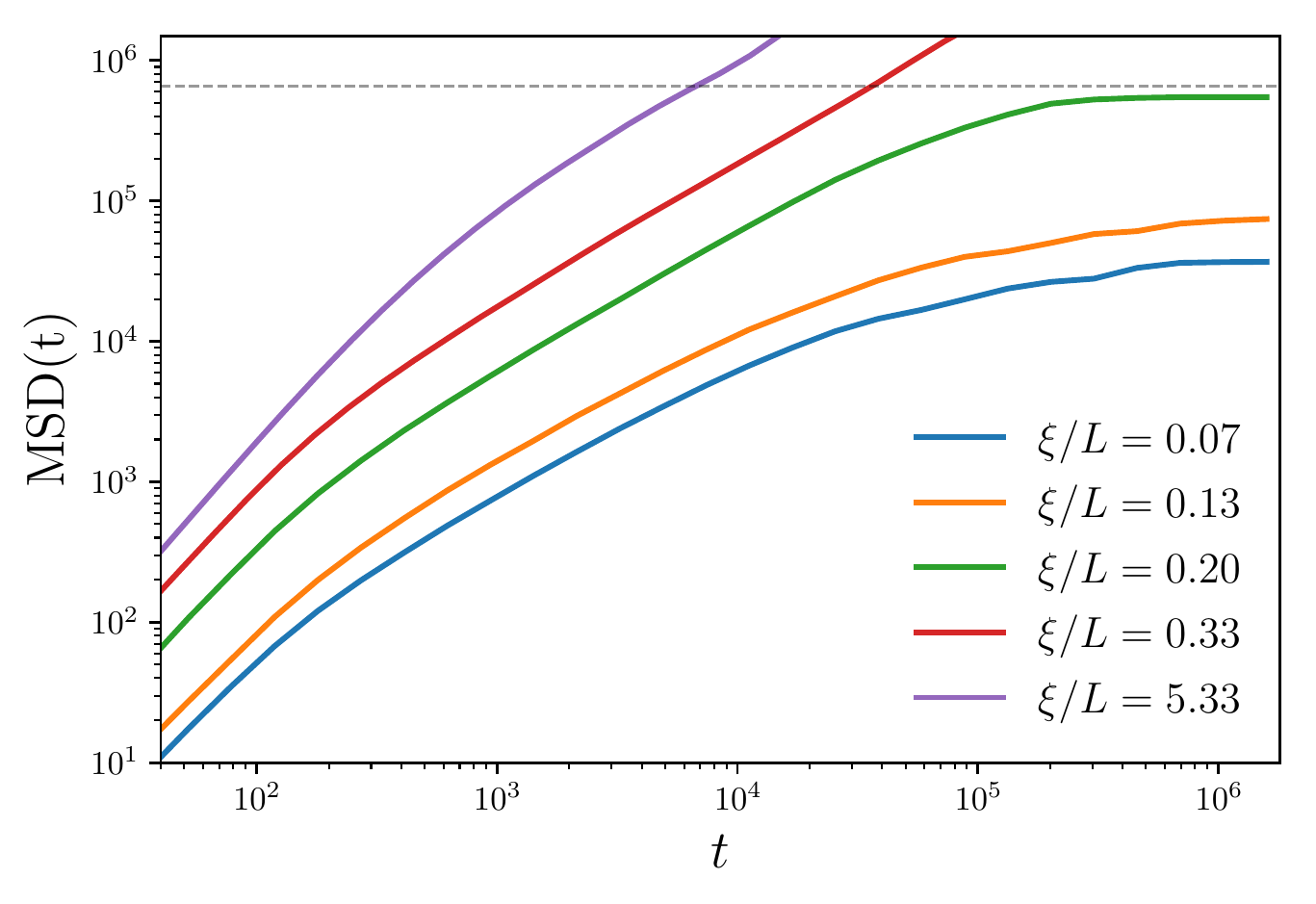} 
		\caption{\small{Mean square displacement of active chains for $\rm{Pe}=945$ and different values of stiffness: from $\xi/L=0.07$ ($K_a=1$) to $\xi/L=5.33$ ($K_a=80$) at $\phi_o=0.6$. The black line indicates the system size.}}
		\label{MSD}	
\end{figure}
\begin{figure}[h!]
  \includegraphics[width=0.16\textwidth,angle=90]{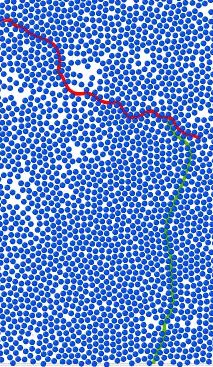}
  \hfill
                \includegraphics[width=0.16\textwidth]{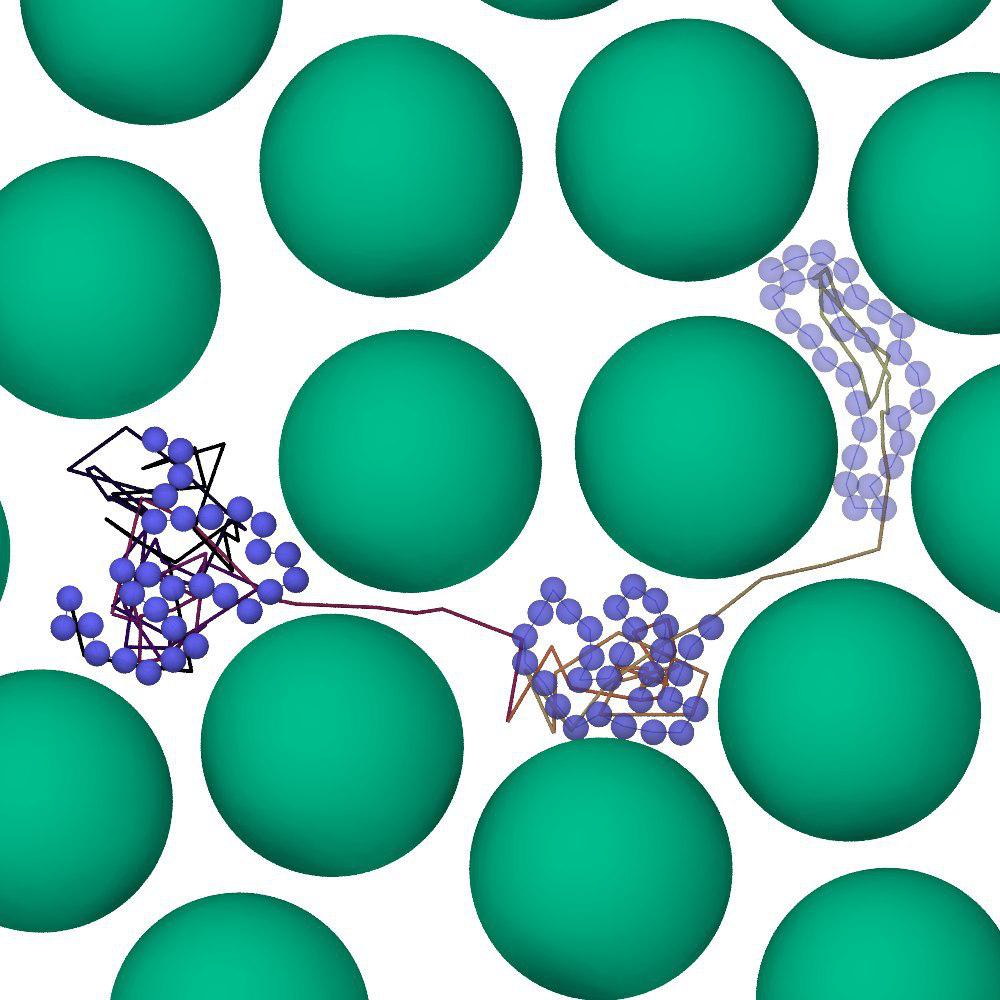}\\
              	\caption{\small{Trajectory $\mb{R}_{CM}(t)$ of stiff
                    filaments ($\xi_p/L=30$)for a time span
                    $\tau\approx 2000$ (left); trajectory of a
                    flexible filament, which curls up and rarely jumps
                    to nearby sites (right).}}
		\label{snapshot-ballistic_stiff_polymers}	
\end{figure}

           
Our focus here is on the dynamics of active polymers in a porous medium.
As a first step we have computed the MSD for various values of the
parameters for filling fraction $\phi=0.6$. 
 An example is shown in Fig.\ref{MSD} for $M=30$
and varying stiffness. For $\xi/L< 0.2$, we observe 3 different regimes:
ballistic due to activity at small $t$, diffusive at intermediate $t$
and saturated or caged for the largest $t$. For larger $\xi/L$, the
polymers are seen to move through the whole sample; for $\xi/L=0.33$, one
can still observe the crossover from ballistic to diffusive motion,
whereas for very large $\xi/L$ the motion is ballistic almost up to system
size. Some trajectories of the very stiff polymers are depicted in
the left panel of Fig.\ref{snapshot-ballistic_stiff_polymers} (see also SI).

\begin{figure}[h!]
		\center	
		\includegraphics[width=0.49\textwidth]{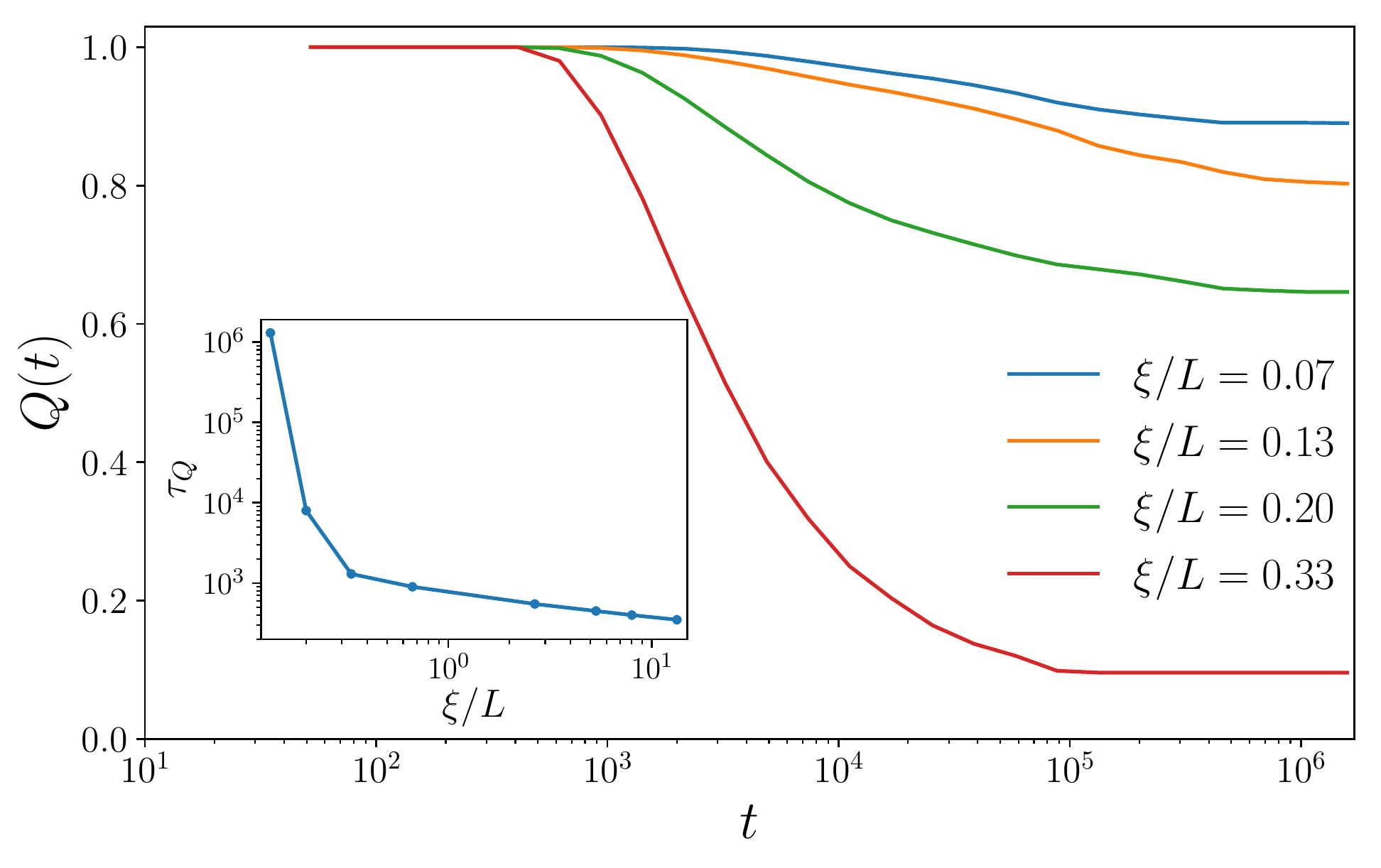} 
		\caption{\small{Fraction of polymers, $Q(t)$, which
                    have moved less than $d=\ell/8$ in time $t$ for
                    several values of persistence length; inset;
                    Relaxation time $\tau$, when Q(t) has decayed to
                    $20\%$ of its initial value.}}
		\label{Qoft}	
\end{figure}

If not all flexible polymers are localised, then the mobile ones will
dominate the MSD for long times. Hence we need a better indicator for
localistion, e.g. the fraction of polymers which have moved less than
a given distance $d$, comparable to the system size. We define
$Q(t)=<\theta(d-|\mb{R}(t_0+t)-\mb{R}(t_0))> $ where the average is
over many different polymers. 
We show $Q(t)$ for $d=\ell/8$ in Fig.(\ref{Qoft}) for the same
data as in Fig.\ref{MSD}. One clearly observes a
time-persistent part for the smallest values of $\xi/L$, indicating that a
finite fraction of the polymers is localised in good agreement with the data from the $\rm{MSD}$. Furthermore the relaxation time of $Q(t)$ grows with decreasing $\xi/L$ and diverges at the critical $\xi$. The inset in Fig.\ref{Qoft} displays the time $\tau_Q$, when more than $20\%$ of the particles have moved by more than $d$. 

 \begin{figure}[h!]
		\center	
		\includegraphics[width=0.49\textwidth]{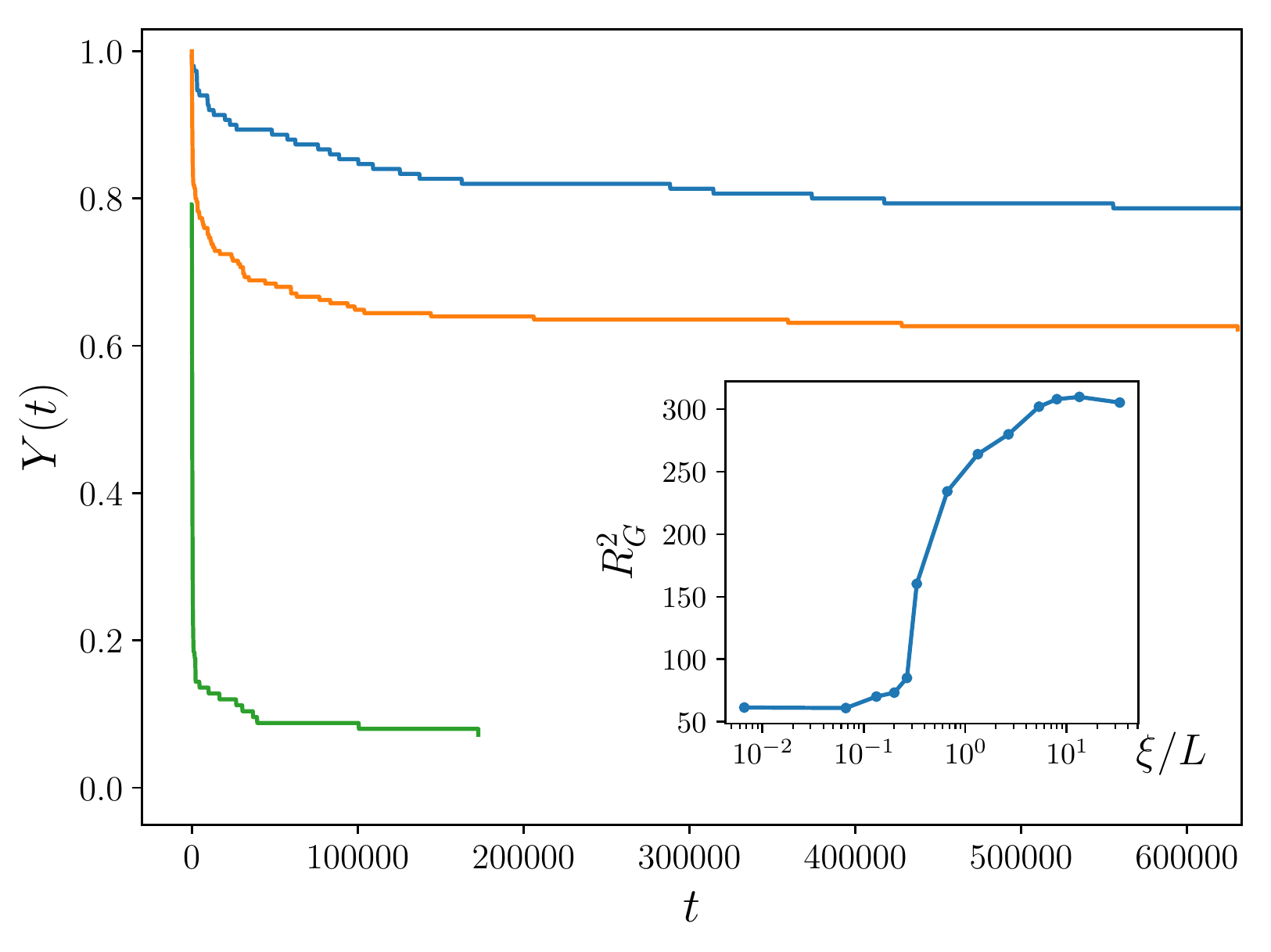} 
		\caption{\small{Fraction of polymers, which have not  at least  unspiraled once in time $t$; parameters $\xi/L=0.07,0.20,0.33$ from top to bottom, $\phi=0.6$, $M=30$; inset: radius of gyration as a function of $\xi/L$; fully extended state corresponds to $R_G\sim 330$. }}
                    		\label{Yoft}	
\end{figure}

Flexible polymers are localised in the porous medium, because they
tend to spiral or fold into a dense state which does not allow them to
pass through the narrow channels. To quantify this statement, we have
computed the radius of gyration for the above set of parameters,
i.e. $\phi=0.6$ and $M=30$. The result is plotted in the inset of
Fig.\ref{Yoft} ; a clear transition is oberved from the dense to the
extended state at around $\xi/L=0.2$.  To pass through the narrow
channels, the polymers have to unspiral, which is a rare event, even
in a system without obstacles and more so in the presence of
obstacles. This can be demonstrated by monitoring the fraction of
polymers which have not unspiralled once in time $t$. We denote this
fraction by $Y(t)$ and plot it for various values of the stiffness in
Fig.\ref{Yoft}. One clearly observes a time persistent part for
$\xi/L \leq 0.2$.

\begin{figure}[h!]
		\center	
		\includegraphics[width=0.49\textwidth]{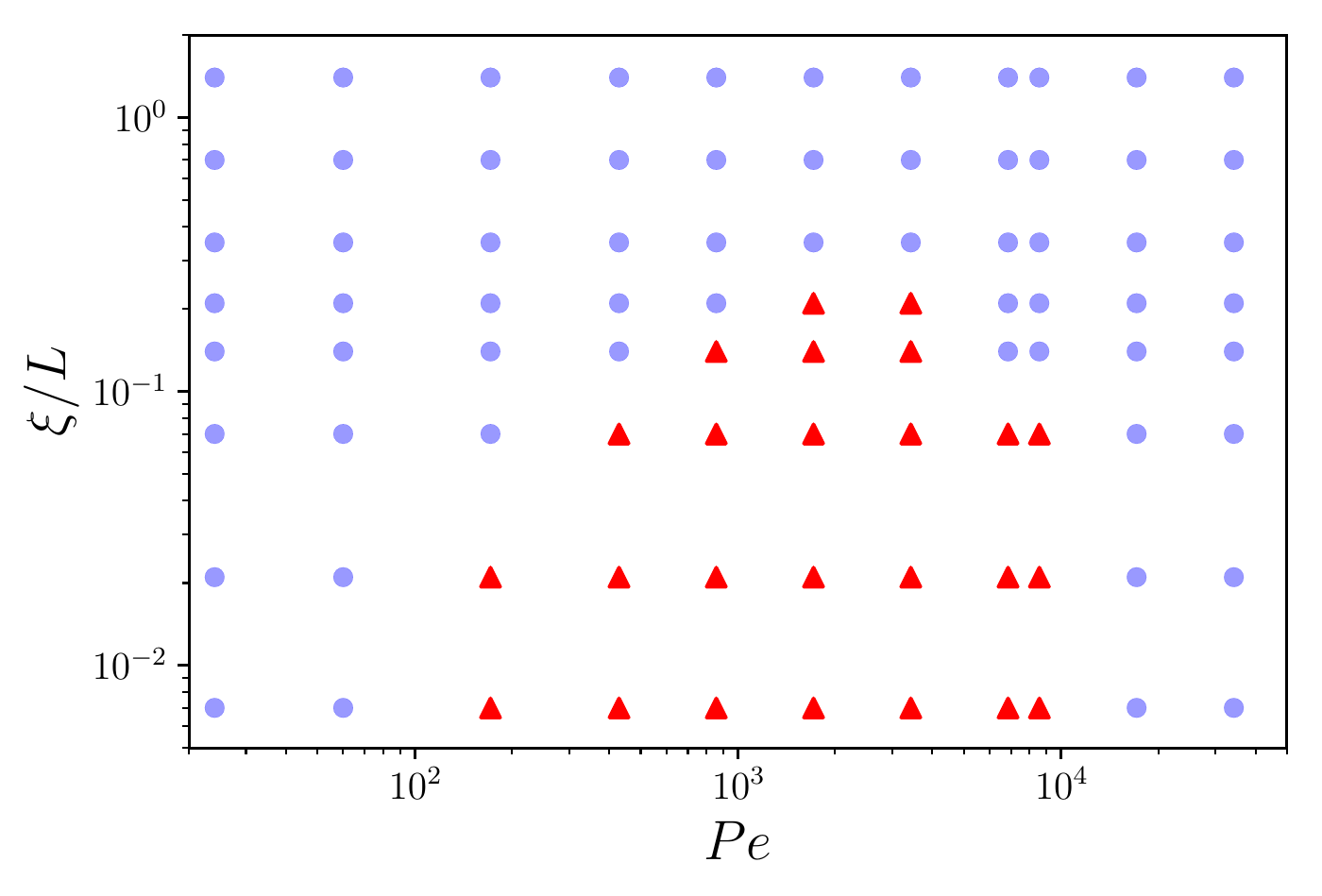} 
		\caption{\small{Phase diagram in the $\xi/L$ and $\rm{Pe}$ plane; parameters $\phi=0.6$ and $M=30$.}}
                    		\label{phase}	
                              \end{figure}

How do these results depend on Peclet number? Spiraling of flexible
polymers is enhanced by high activity or high Peclet number, as
evidenced by the strong spiraling regime in ref.(\cite{isele2015self}). Hence
increasing the Peclet number implies a stronger tendency to
localisation, as displayed in the phase diagram shown in
Fig.\ref{phase}: even stiffer chains get localised with increasing
$\rm{Pe}$.  On the other hand high activity favours directed motion in the
extended state. 
For $\rm{Pe} \geq 3000 $ we observe an increased fraction of
time in the extended state and hence also an increase in the
size of the random steps taken in the extended state, giving rise to
diffusion for the largest $\rm{Pe}$ (see SI). For these very large $\rm{Pe}$, finite size effects become important and possibly shift the phase boundary to larger $\rm{Pe}$.
Localisation also depends on the degree of polymerization,
$M$. Whereas the dynamics of the very stiff ones is hardly effected,
spiraling and hence localisation can only occur for sufficiently long
chains (see SI).

\begin{figure}[h!]
		\center	
		\includegraphics[width=0.49\textwidth]{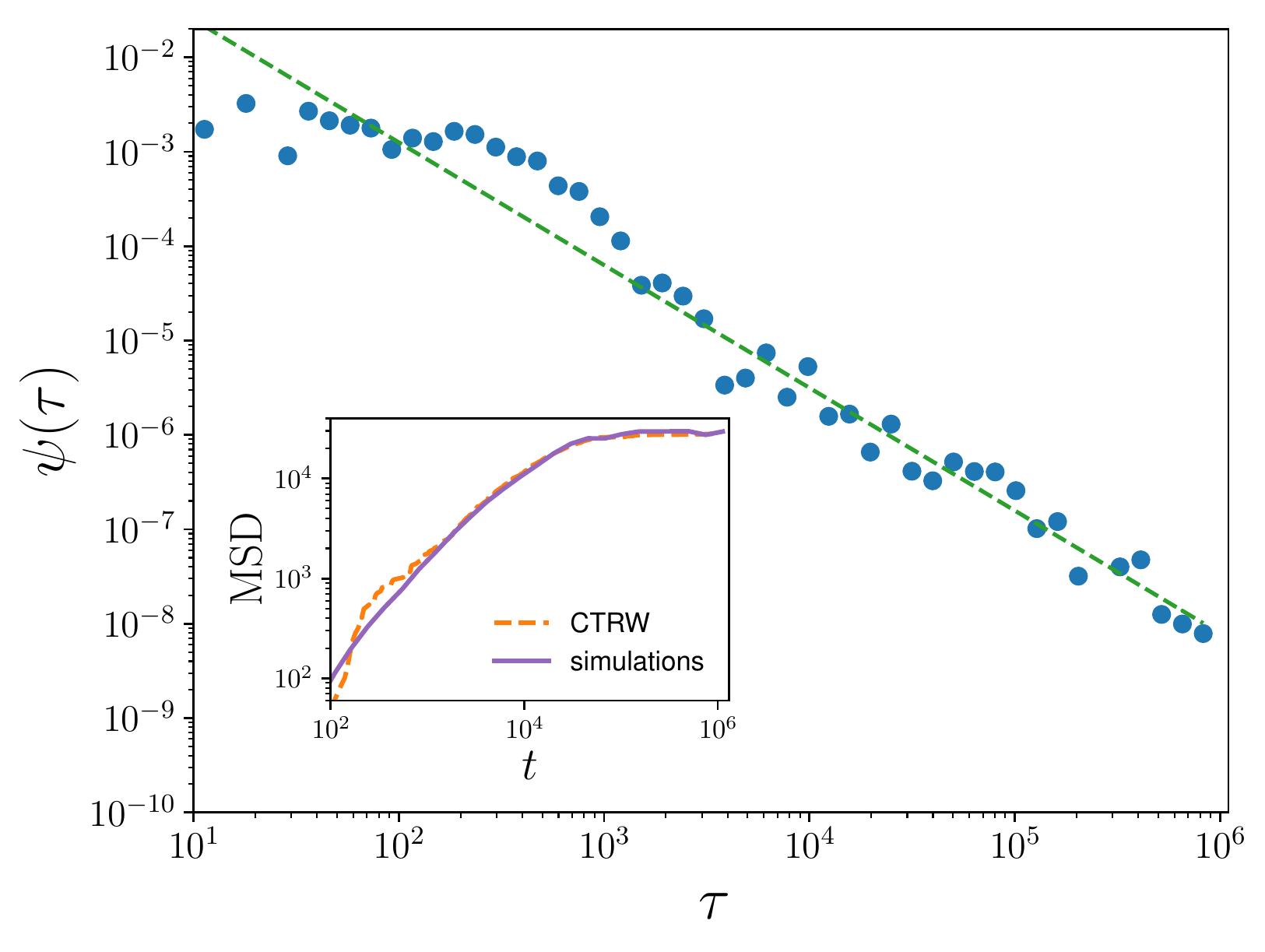} 
		\caption{\small{Waiting time distribution $\Psi(\tau)$
                    for $\xi/L=0.07$ and $\rm{Pe}=945$; inset: comparison of MSD computed
                    from Eq.\ref{MSD_CTRW0} with simulations.}}
                    		\label{Psioft}	
 \end{figure}

  Flexible polymers have to unwind before they can move through the
  narrow channels of the obstructed environment. Our simulations show
  that a flexible chain is most of the time in a dense state,
  unspirals infrequently, makes a sudden jump in the extended state
  and immediately curles up again (see right panel of Fig.\ref{snapshot-ballistic_stiff_polymers} and SI). In a simple model, we neglect the
  time span in the stretched state so that the dynamics can be
  modelled by instanteneous jumps of typical size $\Delta R$,
  separated by random time intervals $\tau.$ 
  The unspiraling of the flexible polymers then corresponds to the
  renewal process of a continuous time random walk
  (CTRW)~\cite{montroll1965random}. The ``renewals'' are the unspiraling
  events, allowing the polymer to make a
  displacment. 
  The central quantity is the distribution of waiting times
  $\psi(\tau)$, where $\tau$ denotes the difference between two
  unspiralling events.
  The mean square displacement
\begin{equation}
\label{MSD_CTRW0}
<(\mb{R}(t)- \mb{R}(0))^2>=<n(t)>\Delta R^2
\end{equation}
is expressed in terms of $<n(t)>$, the mean number of jumps in time $t$. It can be calculated from $\psi(\tau)$:
\begin{equation}
  <n(s)>=\frac{\psi(s)}{s(1-\psi(s))}
\end{equation}
Here we have introduced the Laplace transform of
$n(s)=\int_0^t dt n(t) e^{-st}$ and similarly for $\psi(s)$. The low
frequency behaviour of $\psi(s)$ determines the long time behaviour of
$<n(t)>$ and hence the MSD. If $\psi(s)\sim 1-s<\tau>$ is regular for
samll $s$, then the MSD displays ordinary diffusion. If, on the other
hand, $\psi(s)\sim 1-As^{\alpha}$ with $\alpha<1$, then the MSD is
given by
\begin{equation}
  \label{MSD_CTRW}
  <(\mb{R}(t)- \mb{R}(0))^2>=
  \frac{\Delta R^2 }{A\Gamma(1+\alpha)}\;t^{\alpha}.
\end{equation}
Hence subdiffusive bahaviour should occur and saturation for
$\alpha=0$. 
We have
measured $\psi(\tau)$ for several values of stiffnes $K_a$. An example
is shown in Fig.(\ref{Psioft}). One clearly observes a heavy tail with
an algebraic decay approximately like $t^{-1.3}$ which implies
subdiffusive behaviour $<(\mb{R}(t)- \mb{R}(0))^2> \sim t^{0.3}$ for
the MSD. Due to limited simulation time, we are missing the very
long times, which would lower the exponent to even smaller values. In the inset, we compare the MSD from our simulations to the result of CTRW as given in Eq.\ref{MSD_CTRW0}.

 In conclusion, active polymers in porous media show unexpected, rich
 behaviour.  Whereas stiff chains are able to cross the porous medium
 from one side to the other, flexible chains spiral and are caged.
 Thus, a localisation transition occurs as a function of chain
 stiffness and Peclet number. We have worked out the phase in the
 $\xi/L-\rm{Pe}$ plane and estimated the effects of chain length.
 Localisation of flexible chains occurs due to the persistence of the
 spiraling state. Identifying unspiraling as a renewal event, we can
 model the dynamics of flexible chains as a continuous time random
 walk. The distribution of waiting times between the renewal events was
 shown to develop a heavy tail, as the localisation transition is
 approached, giving rise to subdiffusive behaviour and ultimately
 caging.

 Many extensions of our work lie ahead: Porous media exist in a variety
 of geometries and topologies, giving rise to correspondingly diverse
 dynamics of active polymers. An example is the interplay of pore
 size, filling fraction and chain length in a dense gel. Another
 possible extension are more general models of a crowded environment,
 which is not strictly static but moving slowly as compared to the
 active agent.  Other mechanisms of activity are also of interest,
 such as a dragged chain or the inclusion of a tumbling
 component~\cite{licata2016diffusion}.

 \bibliography{bibfile_literature.bib}{} \bibliographystyle{unsrt}
\end{document}


\title{Supplemental Material: Dynamics of active filaments in porous media}

\author{ Zahra Mokhtari}
\author{Annette Zippelius}
\affiliation{University G\"ottingen, D-37077 G\"ottingen, Germany}
\maketitle

\section*{Movies}
The movies illustrate the qualitative difference between the transport of stiff and flexible active polymers in porous media. To study the single polymer behavior in porous media we have turned off the interaction between different polymers. The obstacle packing fraction is $\phi=0.6$, the Peclet number $\rm{Pe}=945$, and the degree of polymerization $M=30$.
\begin{enumerate}
	\item stiff.chains.mp4 :	\\
	The persistence length relative to the contour length $\xi/L=13$. Stiff active chains traverse large distances almost ballistically and diffuses throughout the whole system at long times.
	\item flexible.chains.mp4\\
	$\xi/L=0.07$. Flexible active chains perform swirling or spiraling inside the pores. They frequently get densely packed between the obstacles in small pores, with their tail beating. They spend long times in the coiled state, rarely unwind and jump to another pore, and wind up in there again.
\end{enumerate}

\section*{Localisation in porous media: role of activity and the degree of polymerization}
\begin{figure}[h!]
		\center	
		\includegraphics[width=0.4\textwidth]{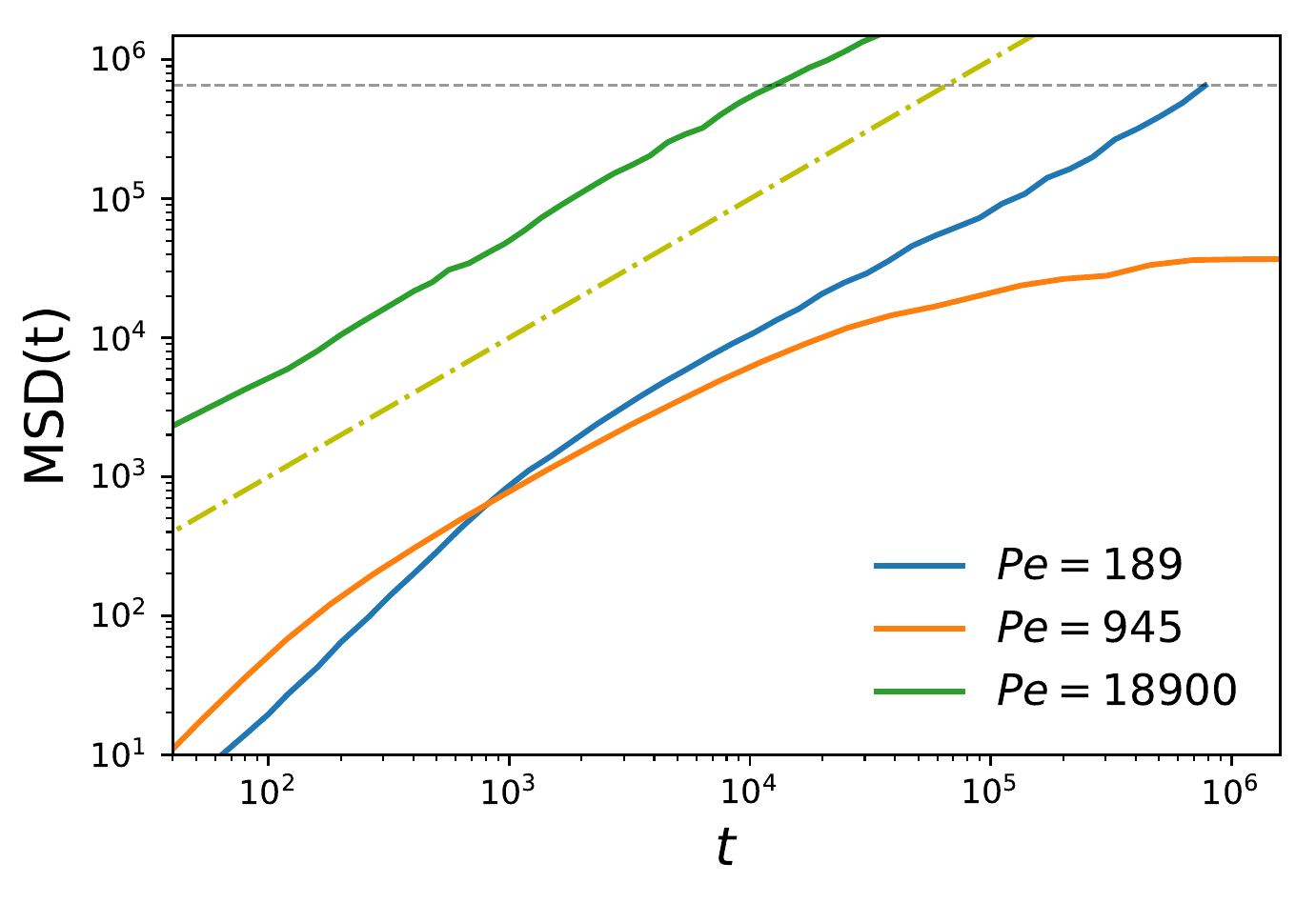} 
		\caption{\small{Mean square displacement of chains with $\xi/L=0.07$, $M=30$, and different $\rm{Pe}$ at $\phi=0.6$. The yellow dashed line illustrates the diffusive regime and the black line is a guide showing the system size.}}
		\label{MSD-diff_pe}	
\end{figure}

\begin{figure}[h!]
		\center	
		\includegraphics[width=0.3\textwidth]{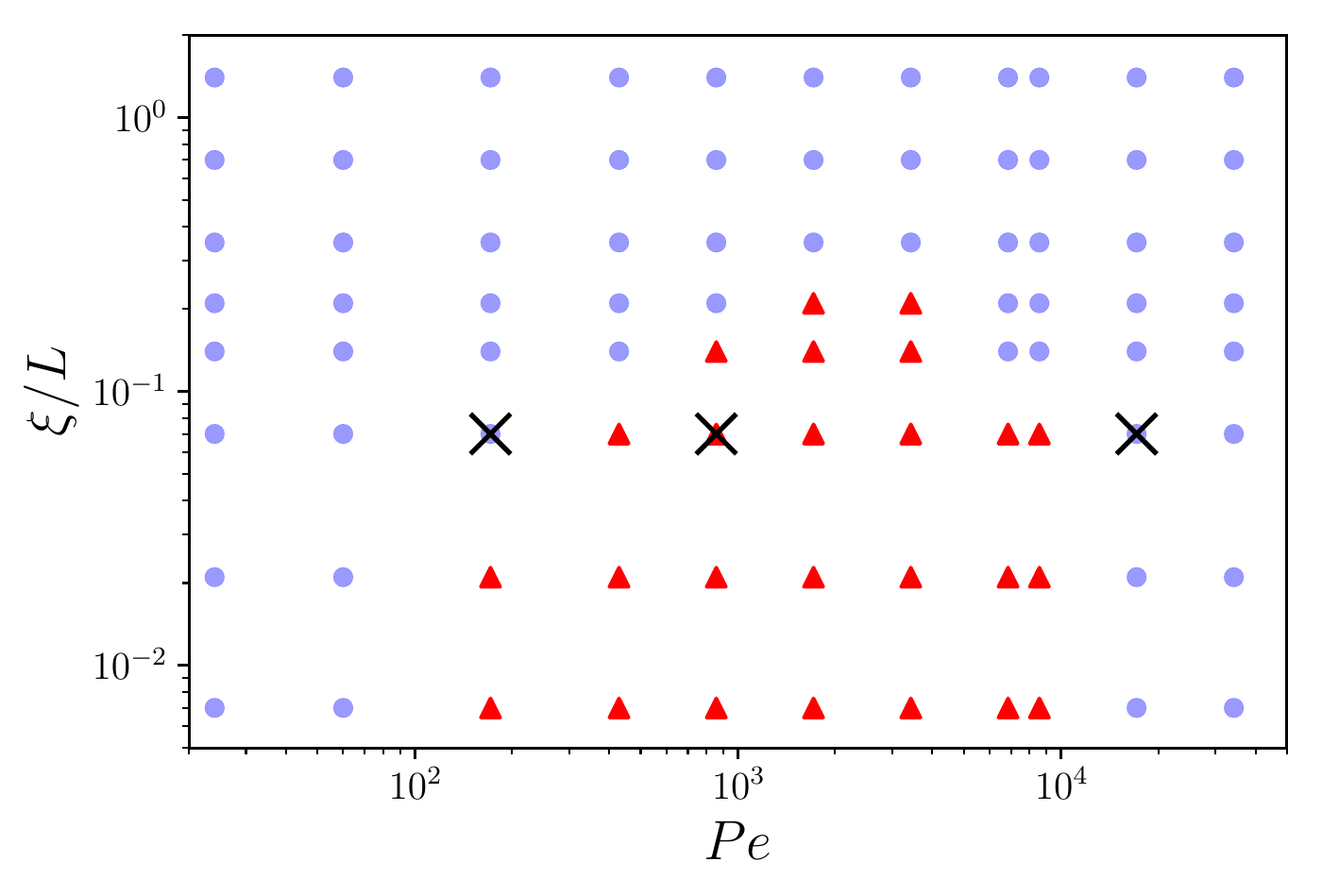} 
		\caption{\small{Phase diagram in the $\xi/L$ and $\rm{Pe}$ plane for $\phi=0.6$ and $M=30$. The three systems in Fig.~\ref{MSD-diff_pe} are indicated with cross signs here.}}
		\label{phase_diagram}	
\end{figure}

Activity enhances the spiraling of flexible polymers~\cite{isele2015self, prathyusha2018dynamically} and hence yields a stronger tendency to localisation. For high activity the fraction of time in the extended state as well as the jump length increase, allowing the polymers to diffuse through the whole system quickly. Fig.~\ref{MSD-diff_pe} illustrates the mean square displacement in 30-bead polymeric systems with $\xi/L=0.07$ and $\phi=0.6$ for different values of $\rm{Pe}$. Localisation is observed at an intermediate activity, below which the swirling is absent, and above which frequent and long jumps result in diffusion. The three chosen values of $\rm{Pe}$ are indicated in the phase diagram with cross signs (Fig.~\ref{phase_diagram}). For huge activities finite size effects become important, yet we have not tracked any localisation in up to 3.2 times larger systems for the largest value of $\rm{Pe}$ in Fig.~\ref{MSD-diff_pe}. 

\begin{figure}[h!]
		\center	
		\includegraphics[width=0.4\textwidth]{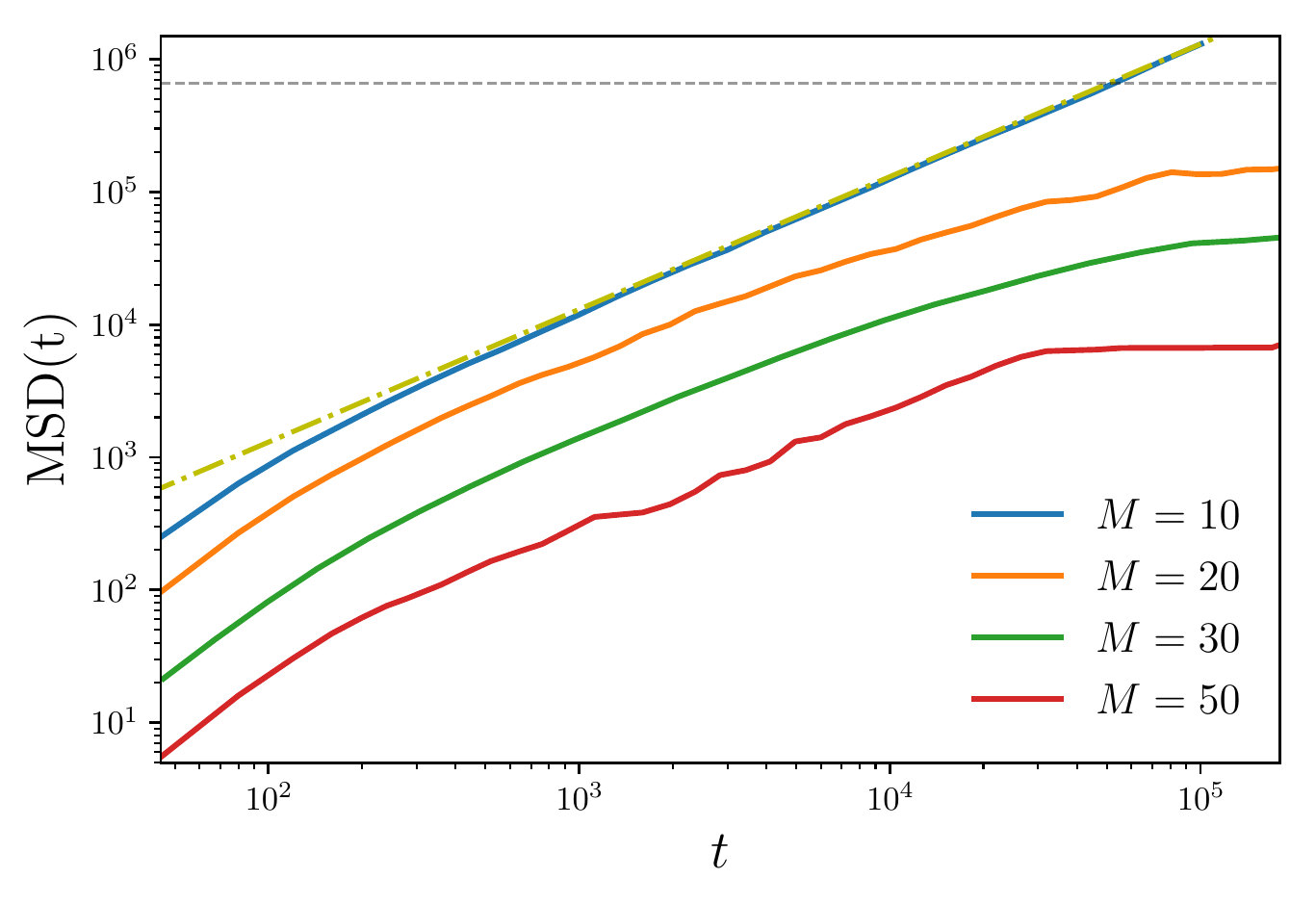} 
		\caption{\small{Mean square displacement of chains with $\xi/L=0.13$, $\rm{Pe}=945$, and different $M$ at $\phi=0.6$. }}
		\label{MSD-diff_M}	
\end{figure}

Spiraling is ruled out for small degree of polymerization. By decreasing $M$ we are moving to the upper left part of the phase diagram and hence weakening the observed localisation phenomenon. Fig.~\ref{MSD-diff_M} shows that shortening the flexible polymers down to $M=10$ eliminates the localisation and allows for diffusion throughout the system. 

\newpage
\bibliography{bibfile_supp.bib}{} \bibliographystyle{plain}